\documentclass[a4paper,onecolumn,unpublished,allowtoday,allowfontchangeintitle,10pt]{quantumarticle}
\pdfoutput=1
\usepackage[utf8]{inputenc}
\usepackage[english]{babel}
\usepackage{amsmath}
\usepackage{hyperref}
\usepackage{mathshortcuts}
\usepackage{mathtools}
\usepackage{amsthm}
\usepackage{cleveref}
\usepackage{sidecap}

\usepackage{tikz}
\usetikzlibrary{decorations,snakes,decorations.markings,positioning,shadows.blur,calc,knots,shapes.misc,shadows,shapes.symbols}
\usepackage{pgfplots}
\pgfplotsset{compat=newest}
\usepgfplotslibrary{fillbetween}
\usepackage{lipsum}

\pgfmathsetmacro{\randamp}{0.005}
\pgfmathtruncatemacro{\totshadow}{30}

\usepackage{multirow}
\usepackage{hhline}
\usepackage{array}
\usepackage{multirow,colortbl}
\newcommand{\PreserveBackslash}[1]{\let\temp=\\#1\let\\=\temp}
\newcolumntype{C}[1]{>{\PreserveBackslash\centering}p{#1}}
\newcolumntype{R}[1]{>{\PreserveBackslash\raggedleft}p{#1}}
\newcolumntype{L}[1]{>{\PreserveBackslash\raggedright}p{#1}}
\makeatletter
\newcommand{\thickhline}{%
	\noalign {\ifnum 0=`}\fi \hrule height 1pt
	\futurelet \reserved@a \@xhline
}
\newcolumntype{"}{@{\hskip\tabcolsep\vrule width 1pt\hskip\tabcolsep}}
\makeatother

\usepackage{glossaries}

\DeclareRobustCommand\antenna{\begin{tikzpicture}[scale=0.5,baseline={([yshift=-1ex]current bounding box.center)}]
		\node at (2,0.25) {\scalebox{0.6}{$\bullet$}};
		\draw (1.8,-0.3) -- (2,0.25) -- (2.2,-0.3);
		\draw (1.8,-0.3) -- (2.135,-0.15) -- (1.91,0) -- (2.04,0.15);
		\draw (2.2,-0.3) -- (1.865,-0.15) -- (2.09,0) -- (1.96,0.15);
		\begin{scope}[xshift=2cm,yshift=0.25cm]
			\draw (140:0.15) arc (140:220:0.15);
			\draw (140:0.25) arc (140:220:0.25);
			\draw (140:0.35) arc (140:220:0.35);
			\draw (40:0.15) arc (40:-40:0.15);
			\draw (40:0.25) arc (40:-40:0.25);
			\draw (40:0.35) arc (40:-40:0.35);
		\end{scope}
\end{tikzpicture}}

\definecolor{LinkColor}{RGB}{167,20,49}
\definecolor{LightGray}{RGB}{220,220,220}
\definecolor{myred}{RGB}{255, 19, 0}
\definecolor{myblue}{RGB}{14, 81, 167}
\definecolor{myorange}{RGB}{255, 129, 0}
\definecolor{mygreen}{RGB}{0, 146, 44}

\hypersetup{
	colorlinks,
	linkcolor=myblue,
	citecolor=myblue,
	filecolor=myblue,
	urlcolor=myblue,
	pdftitle={The debate over QKD: A rebuttal to the NSA's objections},
	pdfauthor={Renato Renner and Ramona Wolf}
}

\theoremstyle{plain}

\theoremstyle{definition}

\crefname{defn}{Definition}{Definitions}
\Crefname{defn}{Definition}{Definitions}
\crefname{thm}{Theorem}{Theorems}
\Crefname{thm}{Theorem}{Theorems}
\crefname{claim}{Claim}{Claims}
\Crefname{claim}{Claim}{Claims}
\crefname{lem}{Lemma}{Lemmas}
\Crefname{lem}{Lemma}{Lemmas}
\crefname{rem}{Remark}{Remarks}
\Crefname{rem}{Remark}{Remarks}
\crefname{prop}{Proposition}{Propositions}
\Crefname{prop}{Proposition}{Propositions}
\crefname{cor}{Corollary}{Corollaries}
\Crefname{cor}{Corollary}{Corollaries}
\crefname{section}{Section}{Sections}
\Crefname{section}{Section}{Sections}
\crefname{equation}{}{}
\Crefname{equation}{}{}
\crefname{figure}{Figure}{Figures}
\Crefname{figure}{Figure}{Figures}
\crefname{appendix}{Appendix}{Appendices}
\Crefname{appendix}{Appendix}{Appendices}
\crefname{table}{Table}{Tables}
\Crefname{table}{Table}{Tables}
\crefname{exmp}{Example}{Examples}
\Crefname{exmp}{Example}{Examples}
\crefname{footnote}{Footnote}{Footnote}
\Crefname{footnote}{Footnote}{Footnote}

\labelcrefformat{subequation}{#2(#1)#3}
\labelcrefrangeformat{subequation}{#3(#1)#4 to #5(#2)#6}

\usepackage{chngcntr}
\counterwithin*{footnote}{section}
\counterwithin{equation}{section}

\allowdisplaybreaks

\makeglossaries
\newglossaryentry{computsecure}
{
	name=computationally secure,
	description={means that security is based on the assumption that a given mathematical problem is hard to solve on a computer.  Cryptosystems that provide this type of security are thus vulnerable to potential future attacks that exploit breakthroughs in software development or novel hardware}
}

\newglossaryentry{inftheosecure}
{
	name=information-theoretically secure,
	description={(also called unconditionally secure) means that security is based on information-theoretic principles. In contrast to a computationally secure cryptographic system, an information-theoretically secure system is immune even to attackers with unlimited computational power}
}

\newglossaryentry{authentication}
{
	name=authentication,
	description={refers to a method for ensuring that the identity of the claimed sender of a message is correct. Authentication requires some initial resources, e.g., a common password held by the sender and the receiver of the message}
}

\newglossaryentry{RSA}
{
	name=RSA,
	description={is a classical algorithm for \gls{publickey} that is nowadays widely used to encrypt data transmission, named after its inventors Rivest, Shamir, and Adleman \cite{Rivest1978}. RSA is \gls{computsecure}, and the problem it relies on is factoring large numbers into prime factors. Because there exists a quantum algorithm for efficiently solving this problem~\cite{Shor1994}, RSA is vulnerable to attackers with access to a (universal) quantum computer}
}

\newglossaryentry{publickey}
{
	name=public-key cryptography,
	description={is a cryptosystem that uses pairs of related keys, consisting of a public and a private key. The public key is openly distributed for others to encrypt data, which can only be decrypted by those who know the corresponding private key. Similarly, public-key cryptography enables other functionalities, such as \gls{authentication} or electronic signatures. Public-key cryptography is usually \gls{computsecure}} 
}

\newglossaryentry{OTP}
{
	name=one-time pad,
	description={(OTP) encryption is a scheme where a message and a \gls{key} are combined via binary addition. The resulting ciphertext does not reveal any information about the encrypted message, but can be decrypted with the same key. This encryption scheme is \gls{inftheosecure}}
}

\newglossaryentry{storenow}
{
	name=``store now decrypt later'' attacks,
	description={exploit the fact that encrypted data can be intercepted during transmission and stored in its encrypted form, to be decrypted once more powerful (quantum) computers are available. These attacks pose a high risk to data that has a long shelf life, like medical records or military secrets}
}

\newglossaryentry{PQC}
{
	name=post-quantum cryptography,
	description={(PQC) refers to classical cryptographic algorithms believed to remain secure even when universal quantum computers are available. The security of these algorithms relies on the assumption that a given mathematical problem is hard to solve for any computational device, including future quantum computers. Post-quantum cryptography is thus usually \gls{computsecure} but not \gls{inftheosecure}}
}

\newglossaryentry{qrepeat}
{
	name=quantum repeaters,
	description={allow to establish entanglement over long distances via a procedure called \emph{entanglement swapping}, effectively enabling the transmission of quantum information over such distances. Since they work entirely on the quantum level, they are secured by the laws of quantum theory and hence don't have to be trusted}
}

\newglossaryentry{sidechannels}
{
	name=side-channel attacks,
	description={do not target the encryption or key distribution protocol itself, but exploit deviations of the implementation from the theoretical description. This could, for example, be leaked information on timing or power consumption, or imperfections of the devices}
}

\newglossaryentry{qmemory}
{
	name=quantum memory,
	description={is the quantum-mechanical analogue of classical computer memory. It stores quantum states for later retrieval}
}

\newglossaryentry{key}
{
	name=cryptographic key,
	description={refers to a bit string that is uniformly random and secret, i.e., known only to the honest communicating parties. This string may then be used, for example, for \gls{OTP} encryption}
}

\newglossaryentry{quantitativeproof}
{
	name=quantitative security proofs,
	description={give a quantitative bound on the probability that a security breach happens (cf.~\cref{fig:quantcost})}
}

\newglossaryentry{protocolsecurity}
{
	name=protocol security,
	description={describes the theoretical security of a protocol}
}

\newglossaryentry{implementationsecurity}
{
	name=implementation security,
	description={denotes the security of a practical implementation of a protocol (which can differ from the theoretical description)}
}

\newglossaryentry{universalqcomputer}
{
	name=universal quantum computers,
	description={are quantum devices that are able to run any quantum algorithm}
}

\begin{document}

\title{The debate over QKD: A rebuttal to the NSA's objections}

\author{Renato Renner}
\email{renner@ethz.ch}
\affiliation{Institute for Theoretical Physics, ETH Zurich, 8093 Zurich, Switzerland}
\affiliation{Quantum Center, ETH Zurich, 8093 Zurich, Switzerland}
\author{Ramona Wolf}
\email{rawolf@phys.ethz.ch}
\affiliation{Institute for Theoretical Physics, ETH Zurich, 8093 Zurich, Switzerland}
\affiliation{Quantum Center, ETH Zurich, 8093 Zurich, Switzerland}


\maketitle

\fontfamily{lmr}\selectfont

\begin{abstract}
A recent publication by the NSA assessing the usability of quantum cryptography has generated significant attention, concluding that this technology is not recommended for use. Here, we reply to this criticism and argue that some of the points raised are unjustified, whereas others are problematic now but can be expected to be resolved in the foreseeable future.

\end{abstract}

\section{Summary of criticism and replies}

The recent publication~\cite{NSAwhitepaper} by the \emph{National Security Agency (NSA)} of the United States assesses the usability and current technical limitations of quantum cryptography and, in particular, quantum key distribution (QKD). It identifies several challenges and concludes that using QKD is not recommended until these challenges are overcome. Similar views have been expressed by the \emph{National Cyber Security Center} of the UK government \cite{NCSC} and the \emph{Agence Nationale de la Sécurité des Systèmes d'Information} of the French government \cite{ANSSI}. While some of the criticism has been addressed in earlier work (see, for example, \cite{Scarani2014,Diamanti2016,NCSCresponse,Alleaume2021}), we provide here specific replies to all points raised in \cite{NSAwhitepaper}. We analyze the limitations raised and discuss to what extent the claims are justified and how QKD can overcome these challenges.\footnote{A short version of these replies has already appeared in \cite{Renner2023}.}

First, we give an overview of the five technical limitations mentioned in \cite{NSAwhitepaper} and present a high-level summary of our replies. Our detailed answers are shown in the next section. The technical terms throughout this note are explained in the~\hyperref[glossary]{Glossary} at the end. The text in italics reproduces the statements in \cite{NSAwhitepaper}. Our assessment of whether these limitations are problematic now, in the medium-term and long-term future, is summarized in \cref{tab:limits}. To avoid providing specific time frames for the terms ``medium-term'' and ``long-term'', we have chosen to define them based on technological milestones---the realization of \gls{qrepeat} and \gls{universalqcomputer}, respectively. This approach is favorable due to the inherent challenge in predicting when these milestones in hardware development will be achieved. By adopting this strategy, we aim to offer an assessment that remains independent of the pace of this development. 

{\renewcommand{\arraystretch}{1.4}\setlength{\tabcolsep}{5pt} 
	\begin{table}[t]
		\centering
		\begin{tabular}{rrC{3cm}C{3cm}C{3cm}}
			\multirow{1}{*}{ } & \multicolumn{1}{c}{ } & \multicolumn{1}{c}{Problematic} & \multicolumn{1}{c}{Problematic} & \multicolumn{1}{c}{Problematic} \\[-0.5em]
			& \multicolumn{1}{c}{ } & \multicolumn{1}{c}{now} & \multicolumn{1}{c}{medium term} & \multicolumn{1}{c}{long term} \\\hhline{=====}
			\multirow{2}{*}{Limitation 1} & \multicolumn{1}{c}{(a)} & \multicolumn{3}{c}{\cellcolor{LightGray!50}not within scope of QKD} \\\hhline{~----}
			 & \multicolumn{1}{c}{(b)} & \multicolumn{1}{C{2.5cm}}{\cellcolor{myorange!20}{see Table~\ref{tab:security}}} & \multicolumn{1}{C{2.5cm}}{\cellcolor{mygreen!20}no} & \multicolumn{1}{C{2.5cm}}{\cellcolor{mygreen!20}no}\\\hhline{-----}
			 \multirow{2}{*}{Limitation 2} & \multicolumn{1}{c}{(a)} & \multicolumn{1}{C{2.5cm}}{\cellcolor{myred!20}yes} & \multicolumn{1}{C{2.5cm}}{\cellcolor{mygreen!20}no} & \multicolumn{1}{C{2.5cm}}{\cellcolor{mygreen!20}no}\\\hhline{~----}
			 & \multicolumn{1}{c}{(b)} & \multicolumn{3}{c}{\cellcolor{LightGray!50}not specific to quantum (vs.\ classical) cryptography}\\\hhline{-----}
			\multirow{2}{*}{Limitation 3} & \multicolumn{1}{c}{(a)} & \multicolumn{1}{C{2.5cm}}{\cellcolor{myred!20}yes} & \multicolumn{1}{C{2.5cm}}{\cellcolor{myorange!20}to some extent} & \multicolumn{1}{C{2.5cm}}{\cellcolor{myorange!20}to some extent}\\\hhline{~----}
			& \multicolumn{1}{c}{(b)} & \multicolumn{1}{C{2.5cm}}{\cellcolor{myred!20}yes} & \multicolumn{1}{C{2.5cm}}{\cellcolor{mygreen!20}no} & \multicolumn{1}{C{2.5cm}}{\cellcolor{mygreen!20}no}\\\hhline{-----}
			\multirow{1}{*}{Limitation 4} & \multicolumn{1}{c}{ } & \multicolumn{1}{C{2.5cm}}{\cellcolor{myred!20}yes} & \multicolumn{1}{C{2.5cm}}{\cellcolor{myred!20}yes} & \multicolumn{1}{C{2.5cm}}{\cellcolor{mygreen!20}no}\\\hhline{-----}
			\multirow{1}{*}{Limitation 5} & \multicolumn{1}{c}{ } & \multicolumn{1}{C{2.5cm}}{\cellcolor{myred!20}yes} & \multicolumn{1}{C{2.5cm}}{\cellcolor{myred!20}yes} & \multicolumn{1}{C{2.5cm}}{\cellcolor{mygreen!20}no}
		\end{tabular}
		\caption{\label{tab:limits}Summary of our assessment of whether Limitations $1$--$5$ are problematic now, in the medium-term, and long-term future. By ``medium-term future'' we mean the epoch when cheaper optical equipment and quantum repeaters are widely available, whereas ``long-term future'' refers to the era when universal quantum computers connected by a quantum network are realized.}
	\end{table}
}

\subsubsection*{Limitation 1: \textit{Quantum key distribution is only a partial solution.}}

\begin{itemize}
	\item[\small(a)] {\small \emph{QKD generates keying material for an encryption algorithm that provides confidentiality. Such keying material could also be used in symmetric key cryptographic algorithms to provide integrity and authentication if one has the cryptographic assurance that the original QKD transmission comes from the desired entity (i.e., entity source authentication). QKD does not provide a means to authenticate the QKD transmission source. Therefore, source authentication requires the use of asymmetric cryptography or preplaced keys to provide that authentication.}}
\end{itemize}

While correct, this statement cannot be regarded as a limitation specific to quantum cryptography. \Gls{authentication} always requires either a pre-shared secret or a trusted third party, independently of whether one uses classical or quantum technology. It is not the goal of QKD to solve this problem.

\begin{itemize}
	\item[\small (b)] {\small \emph{Moreover, the confidentiality services QKD offers can be provided by quantum-resistant cryptography, which is typically less expensive with a better-understood risk profile.}}
\end{itemize}

QKD protocols come with a mathematical proof that they are \gls{inftheosecure}.  Conversely, the security of \gls{PQC} (PQC) protocols---referred to as \emph{quantum-resistant cryptography} above---is only as well understood as that of classical (\gls{computsecure}) schemes. The lack of \gls{quantitativeproof} for the latter is a significant problem, evidenced by a long history of misjudgments. Hence, regarding its \gls{protocolsecurity}, QKD arguably has a better-understood risk profile than PQC  (see also Fig.~\ref{fig:quantcost}). The situation is a bit different if one considers \gls{implementationsecurity} (see~\cref{tab:security}), which is addressed by Limitation~\hyperref[lim:4]{4} (discussed below).

\subsubsection*{Limitation 2: \textit{Quantum key distribution requires special purpose equipment.}}

\begin{itemize}
	\item[\small (a)]{\small\emph{QKD is based on physical properties, and its security derives from unique physical layer communications. This requires users to lease dedicated fiber connections or physically manage free-space transmitters. It cannot be implemented in software or as a service on a network, and cannot be easily integrated into existing network equipment.}}
\end{itemize}

The requirement of dedicated and, thus, expensive hardware is indeed one of the main reasons why QKD is not widely usable today. Nonetheless, such hardware is expected to become more readily available with future advances in optical communication technology. 

\begin{itemize}
	\item[\small (b)]{\small\emph{Since QKD is hardware-based, it also lacks flexibility for upgrades or security patches.}}
\end{itemize}

Any cryptographic scheme, classical or quantum, ultimately runs on hardware, which may be prone to side-channel attacks. The difficulty of patching flawed hardware is thus not a problem specific to quantum cryptography.

\subsubsection*{Limitation 3: \textit{Quantum key distribution increases infrastructure costs and insider threat risks.}}

\begin{itemize}
	\item[]{\small \emph{QKD networks frequently necessitate the use of trusted relays, entailing \vspace{-1mm}
			\begin{itemize}\setlength\itemsep{0mm} 
				\item[\normalfont{(a)}] additional cost for secure facilities and
				\item[\normalfont{(b)}] additional security risk from insider threats.\vspace{-1mm}
			\end{itemize}
	This eliminates many use cases from consideration.}}
\end{itemize}

At the current state, QKD protocols indeed require trusted intermediate stations to achieve longer distances. However, this will change once quantum repeaters are developed. These devices work entirely on the quantum level and hence don't need to be trusted. This eliminates any insider threats and point~(b) will no longer be an issue. Regarding~(a),  while QKD hardware costs are expected to decrease in the coming years, they will likely remain more expensive than classical communication infrastructure.

\subsubsection*{Limitation 4: \textit{Securing and validating quantum key distribution is a significant challenge.}}

\begin{itemize}
	\item[]{\small \emph{The actual security provided by a QKD system is not the theoretical unconditional security from the laws of physics (as modeled and often suggested), but rather the more limited security that can be achieved by hardware and engineering designs. The tolerance for error in cryptographic security, however, is many orders of magnitude smaller than in most physical engineering scenarios making it very difficult to validate. The specific hardware used to perform QKD can introduce vulnerabilities, resulting in several well-publicized attacks on commercial QKD systems.}}
\end{itemize}

The gap between theoretical and implementation security is a general issue in cryptography, already on the classical level. Since quantum communication is a relatively young field, it lacks experience with these problems and is still prone to \gls{sidechannels} (cf.\ Table~\ref{tab:security}). However, this can be resolved by (semi-) device-independent QKD, which requires only minimal (weak) assumptions about the quantum devices and is thus robust against such attacks. Even though this technology is preliminary, it provides a clear path to overcoming this challenge in the long-term future.

\subsubsection*{Limitation 5: \textit{Quantum key distribution increases the risk of denial of service.}}

\begin{itemize}
	\item[]{\small \emph{The sensitivity to an eavesdropper as the theoretical basis for QKD security claims also shows that denial of service is a significant risk for QKD.}}
\end{itemize}

Current implementations of QKD are usually individual point-to-point links. An adversary with access to the link may successfully run a denial-of-service attack. However, future quantum cryptographic solutions are expected to run on a network of quantum connections. Like in classical communication networks, information can be rerouted if one of the links fails to function. Once this stage is reached, there will be no fundamental difference between classical and quantum cryptography regarding their vulnerability to denial-of-service attacks.

\section{Detailed replies}

\subsection{Limitation 1}

\begin{figure}[t]
	\centering
	\begin{tikzpicture}
		\node at (1,1.8) {\textsf{No initial information:}};
		\begin{scope}[xshift=0.04cm,yshift=-0.04cm]
			\path[opacity=0.01,color=myred] foreach \nshadow [evaluate=\nshadow as \angshadow using \nshadow/\totshadow*360] in {1,...,\totshadow}{
				node at (\angshadow:\randamp) {\huge \textsf{A}}
			};
		\end{scope}
		\node[color=myred] at (0,0) {\huge \textsf{A}};
		\begin{scope}[xshift=2.04cm,yshift=0.71cm]
			\path[opacity=0.01,color=gray] foreach \nshadow [evaluate=\nshadow as \angshadow using \nshadow/\totshadow*360] in {1,...,\totshadow}{
				node at (\angshadow:\randamp) {\huge \textsf{B}}
			};
		\end{scope}
		\node[color=gray] at (2,0.75) {\huge \textsf{B}};
		\begin{scope}[xshift=2.04cm,yshift=-0.79cm]
			\path[opacity=0.01,color=gray] foreach \nshadow [evaluate=\nshadow as \angshadow using \nshadow/\totshadow*360] in {1,...,\totshadow}{
				node at (\angshadow:\randamp) {\huge \textsf{E}}
			};
		\end{scope}
		\node[color=gray] at (2,-0.75) {\huge \textsf{E}};
		\draw[decoration={snake},decorate,color=gray,thick] (0.35,0.2) -- (1.65,0.75);
		\draw[decoration={snake},decorate,color=gray,thick] (0.35,-0.2) -- (1.65,-0.75);
		\node[color=black] at (1,0) {\LARGE \textsf{?}};
		\begin{scope}[xshift=4.5cm]
			\node at (1,1.8) {\textsf{Pre-shared secret $s$:}};
			\begin{scope}[xshift=0.04cm,yshift=-0.04cm]
				\path[opacity=0.01,color=myred] foreach \nshadow [evaluate=\nshadow as \angshadow using \nshadow/\totshadow*360] in {1,...,\totshadow}{
					node at (\angshadow:\randamp) {\huge \textsf{A}}
				};
			\end{scope}
			\node[color=myred] at (0,0) {\huge \textsf{A}};
			\begin{scope}[xshift=2.04cm,yshift=0.71cm]
				\path[opacity=0.01,color=mygreen] foreach \nshadow [evaluate=\nshadow as \angshadow using \nshadow/\totshadow*360] in {1,...,\totshadow}{
					node at (\angshadow:\randamp) {\huge \textsf{B}}
				};
			\end{scope}
			\node[color=mygreen] at (2,0.75) {\huge \textsf{B}};
			\begin{scope}[xshift=2.04cm,yshift=-0.79cm]
				\path[opacity=0.01,color=gray] foreach \nshadow [evaluate=\nshadow as \angshadow using \nshadow/\totshadow*360] in {1,...,\totshadow}{
					node at (\angshadow:\randamp) {\huge \textsf{E}}
				};
			\end{scope}
			\node[color=gray] at (2,-0.75) {\huge \textsf{E}};
			\draw[decoration={snake},decorate,color=mygreen,thick] (0.35,0.2) -- (1.65,0.75);
			\node[color=black] (s) at (0.9,0.8) {$s$};
			\draw[color=black,->,>=stealth] (s) to [bend right=10] (0.15,0.35);
			\draw[color=black,->,>=stealth] (s) to [bend left=10] (1.7,1);
			\draw[decoration={snake},decorate,color=gray,thick] (0.35,-0.2) -- (1.65,-0.75);
		\end{scope}
		\begin{scope}[xshift=9cm]
			\node at (1.5,1.8) {\textsf{Trusted third party:}};
			\begin{scope}[xshift=0.04cm,yshift=-0.04cm]
				\path[opacity=0.01,color=myred] foreach \nshadow [evaluate=\nshadow as \angshadow using \nshadow/\totshadow*360] in {1,...,\totshadow}{
					node at (\angshadow:\randamp) {\huge \textsf{A}}
				};
			\end{scope}
			\node[color=myred] at (0,0) {\huge \textsf{A}};
			\draw[fill=myorange!15,draw=myorange] (2,0.75) circle(0.385cm);
			\node[color=myorange] at (2.45,1.2) {$\checkmark$};
			\begin{scope}[xshift=2.04cm,yshift=0.71cm]
				\path[opacity=0.01,color=mygreen] foreach \nshadow [evaluate=\nshadow as \angshadow using \nshadow/\totshadow*360] in {1,...,\totshadow}{
					node at (\angshadow:\randamp) {\huge \textsf{B}}
				};
			\end{scope}
			\node[color=mygreen] (b) at (2,0.75) {\huge \textsf{B}};
			\begin{scope}[xshift=2.04cm,yshift=-0.79cm]
				\path[opacity=0.01,color=gray] foreach \nshadow [evaluate=\nshadow as \angshadow using \nshadow/\totshadow*360] in {1,...,\totshadow}{
					node at (\angshadow:\randamp) {\huge \textsf{E}}
				};
			\end{scope}
			\node[color=gray] at (2,-0.75) {\huge \textsf{E}};
			\draw[decoration={snake},decorate,color=mygreen,thick] (0.35,0.2) -- (1.65,0.75);
			\draw[decoration={snake},decorate,color=gray,thick] (0.35,-0.2) -- (1.65,-0.75);
			\begin{scope}[xshift=3.04cm,yshift=-0.04cm]
				\path[opacity=0.01,color=myorange] foreach \nshadow [evaluate=\nshadow as \angshadow using \nshadow/\totshadow*360] in {1,...,\totshadow}{
					node at (\angshadow:\randamp) {\huge \textsf{T}}
				};
			\end{scope}
			\node[color=myorange] (c) at (3,0) {\huge \textsf{T}};
			\draw[->,>=stealth,color=myorange] (2.95,0.35) to [bend right=10] (2.5,0.75);
			\draw[thick,color=orange] (0.35,0) -- (2.75,0);
			\draw[draw=myorange,fill=myorange!15] (1.6,0) circle (0.2cm);
			\node[color=myorange] at (1.6,0) {$\checkmark$};
		\end{scope}
	\end{tikzpicture}
	\caption{\label{fig:auth}\textbf{Authentication.} Without any initial information about Bob (\textsf{B}), there is no way for Alice (\textsf{A}) to distinguish whether a message she receives is from him or from an adversary Eve (\textsf{E}) who pretends to be~\textsf{B} (left figure). Any authentication scheme (classical or quantum) must rely on something that breaks the symmetry between \textsf{B} and \textsf{E} (from \textsf{A}'s viewpoint). This could be a pre-shared secret $s$ held by \textsf{A} and \textsf{B} (middle figure). Alternatively, \textsf{A} could rely on a trusted third party (\textsf{T}) who can distinguish \textsf{B} from \textsf{E}, which requires some initial authenticated connection between \textsf{A} and \textsf{T} (right figure).}
\end{figure}
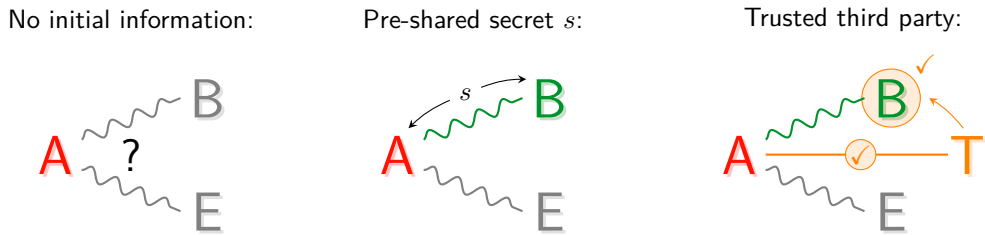

\subsubsection*{Limitation 1\,(a)}
\label{lim:1}

This limitation concerns the fact that QKD does not provide authentication but instead relies on an already established authentic classical communication channel.

\smallskip

Authentication has its price: A party \textsf{A} who wishes to ensure that a message originates from a party~\textsf{B} must either have a pre-shared secret from \textsf{B} or invoke a trusted third party (TTP) who identifies~\textsf{B} towards~\textsf{A}, as illustrated in \cref{fig:auth}. (We refer to Chapter~21 of~\cite{CryptoBook} for a description of how a TTP can facilitate authentication between parties with no prior relationship.) This price must be paid, independently of whether one uses classical or quantum cryptographic protocols. Hence, the reliance of QKD on authenticated communication is not a problem specific to quantum cryptography. 

Crucially, the need for authentication does not compromise the information-theoretic security QKD provides. It has been shown that a small initial secret (for example, a password) shared by \textsf{A} and \textsf{B} is sufficient to establish authentication between them that is information-theoretically secure~\cite{Renner2004,Dodis2009}. Alternatively, if \textsf{A} and \textsf{B} invoke a TTP, information-theoretically secure authentication between them can be established whenever \textsf{A} and \textsf{B}'s initial link to the TTP is information-theoretically secure.\footnote{If \textsf{A} and \textsf{B} each have an authenticated and secure link with the TTP, this can be used to equip \textsf{A} and \textsf{B} with a small shared secret $s$. Afterwards, the two parties are again in the middle scenario of \cref{fig:auth}, in which case information-theoretic authentication is possible as argued before.}

Even if the authentication method used in QKD is not information-theoretically secure but instead relies on (computationally secure) asymmetric cryptography, QKD remains future-proof in the sense that \gls{storenow} do not work. An attacker would have to hack the authentication procedure in real-time to gain access to the generated key. Merely storing the messages exchanged and waiting for more powerful computers to decrypt them would not be sufficient to obtain the key. Once the key generation process is finished, even a complete breach of the authentication procedure does not reveal any information on the generated key.

\subsubsection*{Limitation 1\,(b)}
\label{lim:1b}

In this part of the criticism, it is argued that the confidentiality provided by QKD may as well be achieved with post-quantum cryptography (PQC) \cite{Bernstein2017} (also known as quantum-resistant or quantum-safe cryptography), and it is claimed that the latter has a better-understood risk profile.

\smallskip

To be able to compare the risk profiles of quantum cryptography and PQC, we distinguish two aspects (see \cref{tab:security}):
	\begin{enumerate}
		\item \emph{Protocol security}, which describes the theoretical security of the protocol. 
		\item \emph{Implementation security}, which describes the security of the actual implementation of a protocol (which can deviate from the theoretical description.)
	\end{enumerate}

{\renewcommand{\arraystretch}{1.4}\setlength{\tabcolsep}{5pt} 
	\begin{table}[t]
		\centering
		\begin{tabular}{rrC{2.7cm}C{2.7cm}C{2.7cm}}
			& \multicolumn{1}{c}{ } & \multicolumn{1}{c}{now} & \multicolumn{1}{c}{medium term} & \multicolumn{1}{c}{long term} \\\hhline{=====}
			\multirow{2}{*}{PQC} & \multicolumn{1}{r}{Protocol security} & \multicolumn{1}{C{2.5cm}}{\cellcolor{myred!20}bad} & \multicolumn{1}{C{2.5cm}}{\cellcolor{myred!20}bad} & \multicolumn{1}{C{2.5cm}}{\cellcolor{myred!20}bad}\\\hhline{~----}
			& \multicolumn{1}{r}{Implementation security} & \multicolumn{1}{C{2.5cm}}{\cellcolor{myorange!20}reasonably good} & \multicolumn{1}{C{2.5cm}}{\cellcolor{myorange!20}reasonably good} & \multicolumn{1}{C{2.5cm}}{\cellcolor{myorange!20}reasonably good}\\\hhline{-----}
			\multirow{2}{*}{QKD} & \multicolumn{1}{r}{Protocol security} & \multicolumn{1}{C{2.5cm}}{\cellcolor{mygreen!20}good} & \multicolumn{1}{C{2.5cm}}{\cellcolor{mygreen!20}good} & \multicolumn{1}{C{2.5cm}}{\cellcolor{mygreen!20}good}\\\hhline{~----}
			& \multicolumn{1}{r}{Implementation security} & \multicolumn{1}{C{2.5cm}}{\cellcolor{myred!20}bad} & \multicolumn{1}{C{2.5cm}}{\cellcolor{myorange!20}increasing} & \multicolumn{1}{C{2.5cm}}{\cellcolor{mygreen!20}good}
		\end{tabular}
		\caption{\label{tab:security}Comparison of how well protocol security and implementation security of post-quantum cryptography (PQC) and quantum key distribution (QKD) is understood. Protocol security refers to the abstract protocol. For classical protocols, it usually relies on the conjectured hardness of certain mathematical problems, such as factoring, which is difficult to quantify. Conversely, in quantum cryptography, protocol security relies on physical laws. Implementation security depends on the safety of the hardware and software on which the abstract protocols are run, such as their robustness against side-channel attacks. Here classical cryptography has an advantage compared to quantum cryptography due to the experience acquired over many decades, whereas quantum hardware and software engineering is still in the early stages.}
	\end{table}

The protocol security of PQC relies on the assumption that a given mathematical problem is hard to solve for classical and quantum computers. The crux is that evidence for such an assumption is sparse. It depends on how many mathematicians or computer scientists have already tried to solve the problem and for how long. The list of problems considered ``hard'' is thus generally shrinking over time (\cref{fig:quantcost}). Furthermore, while researchers have decade-long experience regarding hard problems for classical computers, quantum computing is relatively young, and it is conceivable that novel quantum algorithms for solving problems that were initially considered hard will be discovered (as was already the case for the factoring problem).

A PQC protocol may thus turn insecure overnight. This is not merely a theoretical concern but a practically relevant threat, as evidenced in the context of the standardization process for PQC of the National Institute of Standards and Technology (NIST) \cite{NIST}. This search spanned several years until some finalists for standardization, as well as some alternatives, were announced in 2022. However, it only took a couple of months until one of the alternatives, called SIKE (which is short for supersingular isogeny key encapsulation), was broken on a single-core classical computer \cite{SIKE2022}. 

The history of cryptography is full of other examples that illustrate the difficulty of assessing and quantifying protocol security of computational cryptography in general. For instance, the inventors of the widely-used \gls{RSA} encryption scheme initially calculated that factoring a 200-digit number would take a few billion years with the best-known factoring method \cite[Table 1]{Rivest1978}, which is on the order of the estimated remaining lifespan of the universe. As a result, they recommended the use of 200-digit keys. Nonetheless, in 2020, a 250-digit number was factored \cite{RSA250}.

Conversely, the protocol security of QKD is provable based on the laws of physics. It is thus unaffected by algorithmic discoveries or hardware developments.  In addition, the protocol security can be quantified in terms of a bound on the probability that the protocol is broken. The risk profile of QKD protocols is thus perfectly understood (see \cref{fig:quantcost}). Hence, regarding protocol security, QKD has a clear advantage compared to PQC.  

PQC, however, has an advantage compared to QKD in terms of implementation security, although this advantage is temporary. Implementations of PQC can draw on decades of experience with classical computers, which has led to a good understanding of potential side-channel attacks.  On the other hand, the implementation security of QKD is still in the exploratory stage. As QKD is a relatively young technology, researchers have only little experience with possible side-channel attacks and countermeasures. Still, this understanding will increase in the coming years. Furthermore, in the medium and long term future, the issue can be resolved with semi-device-independent and fully device-independent QKD, respectively (see the discussion of Limitation~\hyperref[lim:4]{4}). 

\begin{figure}[t]
	\centering
	\begin{tikzpicture}[scale=1.05]
		\draw[thick,->,>=stealth] (0,0) -- (9.25,0);
		\draw[thick,->,>=stealth] (0,0) -- (0,4);
		\node at (9.65,0) {\footnotesize time};
		\node at (0,4.45) {\footnotesize probability of};
		\node at (0,4.2) {\footnotesize security breach};
		\draw[thick, color=myred] (-0.15,0.25) -- (0,0.25) -- (1.5,0.5) -- (1.5,0.75) -- (3,1) -- (3,1.3) -- (4,1.5) -- (4,3.5) -- (9,3.5);
		\draw[thick, color=myblue] (-0.15,0.25) -- (0,0.25) -- (1.25,0.3) -- (1.25,0.5) -- (2.5,0.75) -- (2.5,1.25) -- (3.75,1.5) -- (3.75,1.8) -- (5,2) -- (5,2.3) -- (6.5,2.5) -- (6.5,2.7) -- (7.25,2.9) -- (7.25,3.15) -- (9,3.45);
		\draw[thick,color=mygreen] (-0.15,0.25) -- (9,0.25);
		\node[color=myblue] at (8.5,3) {\footnotesize post-quantum};
		\node[color=myblue] at (8.5,2.75) {\footnotesize protocol};
		\node[color=mygreen] at (8,0.5) {\footnotesize QKD protocol};
		\node[color=myred] at (8,4) {\footnotesize RSA};
		\node[color=myred] at (8,3.75) {\footnotesize protocol};
		\draw[color=black,dashed] (4,-0.15) -- (4,1.5);
		\node[color=black] at (4,-0.4) {\footnotesize first universal};
		\node[color=black] at (4,-0.65) {\footnotesize quantum};
		\node[color=black] at (4,-0.9) {\footnotesize computer};
		\node[color=black] at (0.8,1.5) {\footnotesize algorithmic};
		\node[color=black] at (0.8,1.25) {\footnotesize discovery};
		\draw[->,>=stealth] (1.35,1.2) to[bend left] (1.5,0.85);
		\draw[->,>=stealth] (1.5,1.4) to[bend left] (2.4,1.3);
		\node at (2,2.3) {\footnotesize evolution of};
		\node at (2,2.05) {\footnotesize hardware};
		\draw[->,>=stealth] (2.65,2) to[bend left] (3.25,1.55);
		\draw[->,>=stealth] (2.8,2.25) to[bend left] (4.25,2);
	\end{tikzpicture}
	\caption{\label{fig:quantcost}\textbf{Protocol security of cryptographic protocols over time.} The diagram shows schematically the development of the probability that an encryption protocol is broken if the adversary has all the computational power in the world as a function of time. Classical protocols (including post-quantum ones), which aim to provide computational security, are becoming increasingly insecure due to the evolution of hardware and algorithmic discoveries. If an efficient quantum algorithm is found for breaking it (which is the case for RSA), the scheme will become insecure once the first universal quantum computer is available. On the other hand, the failure probability of quantum key distribution always remains the same because it only relies on the laws of quantum physics, which don't change over time. This figure is taken from \cite{Renner2023}.}
\end{figure}
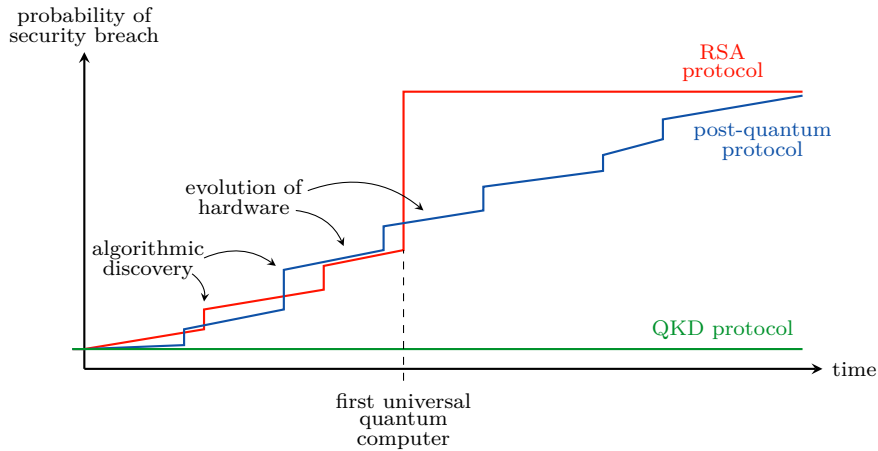

\subsection{Limitation 2}
\label{lim:2}

This limitation concerns the need for dedicated hardware when implementing QKD. 

\subsubsection*{Limitation 2\,(a)}
\label{lim:2a}

The first part of this criticism is that QKD cannot be easily integrated into existing network equipment.  

\smallskip

QKD requires a communication link that transports information encoded into one single quantum optical mode from sender to receiver at high fidelity. In today's implementations of quantum cryptography, this is realized by point-to-point optical fiber or free-space connections. Current optical communication networks, however, do not provide such high-fidelity links. Hence, integrating QKD indeed requires expensive special-purpose hardware. 

Nonetheless, the steady improvements in the efficiency of classical optical communication are expected to eventually reach a point where one (or even more) bits are encoded per photon \cite{Banaszek2020}. In this way, classical technology will naturally approach the requirements for quantum communication, thus facilitating a more straightforward and cheaper integration of QKD. 

\subsubsection*{Limitation 2\,(b)}

The second part of Limitation~2 refers to the difficulty of administering security patches. 

\smallskip

Here it is again helpful to distinguish protocol and implementation security as in the discussion of Limitation~\hyperref[lim:1b]{1\,(b)}. Since QKD comes with a mathematical proof of security, the protocol parameters do not require any updates. This is different in computational cryptography, where algorithmic or hardware breakthroughs may imply that security parameters, such as the key length of RSA~\cite{Barker2015}, need to be adapted. 

Regarding implementation security, there is no fundamental difference between classical and quantum cryptography. If hardware is found to be flawed or prone to side-channel attacks, patches on the hardware level will be required in both cases.

\subsection{Limitation 3}
\label{lim:3}

This limitation derives from the claim that QKD networks require trusted relays. 

\smallskip

It is correct that current implementations of QKD require trusted relays. The typical information carrier in quantum communication is single photons. Since the loss of photons in optical fibers is typically very high, at the moment, intermediate stations are required to achieve large distances \cite{Huttner2022}.
The problem of signal losses arises in classical communication, too, requiring the use of repeaters. They measure the incoming signal, copy it, and retransmit it to the other side at higher power, thus effectively amplifying the signal. But the same technique does not work for quantum information, because copying it is prohibited fundamentally, as asserted by the no-cloning theorem \cite{Wootters1982,Dieks1982}. Therefore, current implementations of QKD are restricted to point-to-point connections that have no repeaters in between. When combining such point-to-point connections to form a network, the communication must be encoded and decoded separately for any link of the network (see \cref{fig:links}). But since the intermediate nodes need to store secret \emph{classical} information, they must be trusted. 

However, the need for trusted relays is not fundamental---quantum repeaters \cite{Briegel1998,Sangouard2011,Azuma2022} will replace them in the medium-term future. Quantum repeaters work coherently on the quantum level and are thus secured by the laws of quantum theory in the same way as QKD is secured by these laws. Hence, even if they are hacked and controlled by a quantum adversary, security is still guaranteed. While this method is well-established in theory, it has yet to be experimentally realized. The main obstacle is that a quantum repeater requires \gls{qmemory}. The storage time of state-of-the-art quantum memories is insufficient to outperform direct optical links, despite considerable progress in recent years \cite{Liu2021,LagoRivera2021,Rabbie2022}. However, since quantum memories are a crucial part of quantum computers, they are being intensively researched on various technology platforms.

\subsubsection*{Limitation 3\,(a)}
\label{lim:3a}

This part of the criticism concerns the costs for trusted relays. 

\smallskip

As explained above, current implementations of QKD require trusted relays. These must be placed in secure facilities, which are costly. 

In the medium-term future, the trusted relays can be replaced by quantum repeaters. While it is expected that these devices become cheaper as optical technology develops (see also Limitation~\hyperref[lim:2a]{2\,(a)}), building a quantum communication network will most likely remain more expensive than the corresponding infrastructure for classical communication.

\begin{figure}[t]
	\centering
	\begin{tikzpicture}
		\begin{scope}[xshift=0.04cm,yshift=-0.04cm]
			\path[opacity=0.01,color=myred] foreach \nshadow [evaluate=\nshadow as \angshadow using \nshadow/\totshadow*360] in {1,...,\totshadow}{
				node at (\angshadow:\randamp) {\huge \textsf{A}}
			};
		\end{scope}
		\node[color=myred] at (0,0) {\huge \textsf{A}};
		\begin{scope}[xshift=8.04cm,yshift=-0.04cm]
			\path[opacity=0.01,color=myblue] foreach \nshadow [evaluate=\nshadow as \angshadow using \nshadow/\totshadow*360] in {1,...,\totshadow}{
				node at (\angshadow:\randamp) {\huge \textsf{B}}
			};
		\end{scope}
		\node[color=myblue] at (8,0) {\huge \textsf{B}};
		\node at (2,0.25) {$\bullet$};
		\draw (1.8,-0.3) -- (2,0.25) -- (2.2,-0.3);
		\draw (1.8,-0.3) -- (2.135,-0.15) -- (1.91,0) -- (2.04,0.15);
		\draw (2.2,-0.3) -- (1.865,-0.15) -- (2.09,0) -- (1.96,0.15);
		\begin{scope}[xshift=2cm,yshift=0.25cm]
			\draw (140:0.15) arc (140:220:0.15);
			\draw (140:0.25) arc (140:220:0.25);
			\draw (140:0.35) arc (140:220:0.35);
			\draw (40:0.15) arc (40:-40:0.15);
			\draw (40:0.25) arc (40:-40:0.25);
			\draw (40:0.35) arc (40:-40:0.35);
		\end{scope}
		\begin{scope}[xshift=2cm]
			\node at (2,0.25) {$\bullet$};
			\draw (1.8,-0.3) -- (2,0.25) -- (2.2,-0.3);
			\draw (1.8,-0.3) -- (2.135,-0.15) -- (1.91,0) -- (2.04,0.15);
			\draw (2.2,-0.3) -- (1.865,-0.15) -- (2.09,0) -- (1.96,0.15);
			\begin{scope}[xshift=2cm,yshift=0.25cm]
				\draw (140:0.15) arc (140:220:0.15);
				\draw (140:0.25) arc (140:220:0.25);
				\draw (140:0.35) arc (140:220:0.35);
				\draw (40:0.15) arc (40:-40:0.15);
				\draw (40:0.25) arc (40:-40:0.25);
				\draw (40:0.35) arc (40:-40:0.35);
			\end{scope}
		\end{scope}
		\begin{scope}[xshift=4cm]
			\node at (2,0.25) {$\bullet$};
			\draw (1.8,-0.3) -- (2,0.25) -- (2.2,-0.3);
			\draw (1.8,-0.3) -- (2.135,-0.15) -- (1.91,0) -- (2.04,0.15);
			\draw (2.2,-0.3) -- (1.865,-0.15) -- (2.09,0) -- (1.96,0.15);
			\begin{scope}[xshift=2cm,yshift=0.25cm]
				\draw (140:0.15) arc (140:220:0.15);
				\draw (140:0.25) arc (140:220:0.25);
				\draw (140:0.35) arc (140:220:0.35);
				\draw (40:0.15) arc (40:-40:0.15);
				\draw (40:0.25) arc (40:-40:0.25);
				\draw (40:0.35) arc (40:-40:0.35);
			\end{scope}
		\end{scope}
		\draw (0.35,0) to node[above] {\small $K_1$} (1.6,0);
		\draw (2.4,0) to node[above] {\small $K_2$} (3.6,0);
		\draw (4.4,0) to node[above] {\small $K_3$} (5.6,0);
		\draw (6.4,0) to node[above] {\small $K_4$} (7.65,0);
		\begin{scope}[yshift=-1.5cm]
			\begin{scope}[xshift=0.04cm,yshift=-0.04cm]
				\path[opacity=0.01,color=myred] foreach \nshadow [evaluate=\nshadow as \angshadow using \nshadow/\totshadow*360] in {1,...,\totshadow}{
					node at (\angshadow:\randamp) {\huge \textsf{A}}
				};
			\end{scope}
			\node[color=myred] at (0,0) {\huge \textsf{A}};
			\begin{scope}[xshift=8.04cm,yshift=-0.04cm]
				\path[opacity=0.01,color=myblue] foreach \nshadow [evaluate=\nshadow as \angshadow using \nshadow/\totshadow*360] in {1,...,\totshadow}{
					node at (\angshadow:\randamp) {\huge \textsf{B}}
				};
			\end{scope}
			\node[color=myblue] at (8,0) {\huge \textsf{B}};
			\draw[fill=LightGray,blur shadow={shadow blur steps=5}] (2,0) circle (0.4cm);
			\node at (2,0) {\small $\mathsf{QR}$};
			\draw[fill=LightGray,blur shadow={shadow blur steps=5}] (4,0) circle (0.4cm);
			\node at (4,0) {\small $\mathsf{QR}$};
			\draw[fill=LightGray,blur shadow={shadow blur steps=5}] (6,0) circle (0.4cm);
			\node at (6,0) {\small $\mathsf{QR}$};
			\draw[decoration={snake},decorate,color=black] (0.35,0) to node[above] {\small $\Psi_1$} (1.6,0);
			\draw[decoration={snake},decorate,color=black] (2.4,0) to node[above] {\small $\Psi_2$} (3.6,0);
			\draw[decoration={snake},decorate,color=black] (4.4,0) to node[above] {\small $\Psi_3$} (5.6,0);
			\draw[decoration={snake},decorate,color=black] (6.4,0) to node[above] {\small $\Psi_4$} (7.65,0);
 		\end{scope}
	\end{tikzpicture}
	\caption{\label{fig:links} \textbf{Long-distance QKD via trusted intermediate stations vs.\ quantum repeaters.} The use of trusted intermediate stations, depicted as \antenna, requires establishing a secret key $K_i$ on each segment. Since these keys are secret classical information, the stations must be trusted. Quantum repeaters ($\mathsf{QR}$), on the other hand, work entirely on the quantum level (illustrated by quantum states $\Psi_i$). <They are hence secured by the laws of quantum theory and don't have to be trusted.}
\end{figure}

\subsubsection*{Limitation 3\,(b)}

This part of the criticism refers to security risks from insider threats. 

\smallskip

Security proofs of QKD extend directly to links with intermediate quantum repeaters. Hence, in the medium-term future, when trusted nodes are replaced by quantum repeaters, there do not occur any additional risks from insider threats.

\subsection{Limitation 4}
\label{lim:4}

This limitation refers to the gap between the security of the theoretical protocol and the security of the practical implementation.\footnote{One should note that, while for post-quantum cryptography the implementation security is generally better understood (see \cref{tab:security}), one still has to watch out for new developments regarding side-channel attacks. A recent example that illustrates this fact comes from advances in artificial intelligence (AI) research. Attacks based on machine learning can analyze large amounts of measurable data obtained from a device running the implementation, such as timing and power consumption, possibly recovering the original message. This was, for example, demonstrated for one of the finalists of the NIST standardization process called CRYSTALS-Kyber \cite{Dubrova2022}. Even though this kind of attack does not break the algorithm itself but is a side-channel attack on the \emph{implementation}, it shows that AI-assisted attacks pose a real threat to the practical security post-quantum and classical cryptography can offer.} 

\smallskip

Devices used within an implementation, such as quantum sources and detectors, often deviate from their theoretical description. This can open up side channel attacks, which exploit such imperfections, both of the quantum source and the detector (see \cite{Makarov2006,Lydersen2010,Gerhardt2011} for some examples). One approach to prohibit such attacks is to adapt the protocols, or the relevant parameters, in such a way that known imperfections can be tolerated (see, for example, \cite{Gottesman2004,Tamaki2014,Pereira2023}). However, the imperfections are often unknown, especially in real-world implementations, where the devices are exposed to changing environmental conditions.

\begin{figure}[t]
	\centering
	\begin{tikzpicture}[scale=1.25]
		\node at (1,0.5) {\small OTP (abstract):};
		\draw (0,0) -- (0.85,0);
		\node at (1,0) {$\bigoplus$};
		\draw (0,-0.75) -- (1,-0.75) -- (1,-0.15);
		\node at (1,-0.75) {$\bullet$};
		\draw (1.15,0) -- (2,0);
		\node at (-0.25,0) {\small $M$};
		\node at (-0.25,-0.75) {\small $K$};
		\node at (2.25,0) {\small $C$}; 
		\begin{scope}[yshift=-2.2cm]
			\node at (1,0.5) {\small OTP (real):};
			\draw (0,0) -- (0.85,0);
			\draw (0,-0.75) -- (1,-0.75) -- (1,-0.15);
			\draw (1.15,-0.375) -- (2,-0.375);
			\draw[fill=gray!30!white] (0.75,-0.9) -- (1.25,-0.9) -- (1.25,0.15) -- (0.75,0.15) -- cycle;
			\node at (1,-0.375) {$\mathsf{+}$};
			\node at (-0.25,0) {\small $V_M$};
			\node at (-0.25,-0.75) {\small $V_K$};
			\node at (2.25,-0.375) {\small $V_C$}; 
			\draw[fill=myblue!15,draw=myblue] (1.6,-1.2) -- (2,-1.2) -- (2,-0.8) -- (1.6,-0.8) -- cycle;
			\node[color=myblue] at (1.8,-1) {\small $\mathsf{S}$};
			\draw[->,>=stealth,draw=myblue,thick] (1.25,-0.375) to [bend left=20] (1.75,-0.75);
			\draw[->,>=stealth,draw=myblue,thick] (1.85,-0.75) to [bend left=20] (2.15,-0.5);
		\end{scope}
		\begin{scope}[xshift=4.5cm,yshift=-1cm,scale=1.1]
			\draw[fill=myblue!15,draw=none] (0,-1) -- (2.4,-1) -- (2.4,-0.5) -- (1.8,-0.5) -- (1.8,0.3) -- (1.2,0.3) -- (1.2,0.4) -- (0.6,0.4) -- (0.6,-0.6) -- (0,-0.6) -- cycle;
			\node at (0,1.25) {\small $V_C$};
			\draw[color=gray] (0.6,-1.9) -- (0.6,0.9);
			\draw[color=gray] (1.2,-1.9) -- (1.2,0.9);
			\draw[color=gray] (1.8,-1.9) -- (1.8,0.9);
			\node at (0.3,-1.25) {\footnotesize $M\!\!=\!\!0$};
			\node at (0.3,-1.5) {\footnotesize $K\!\!=\!\!0$};
			\node[color=myblue] at (0.3,-1.75) {\footnotesize $C\!\!=\!\!0$};
			\draw[color=myblue,thick] (0,-0.6) -- (0.6,-0.6);
			\node at (0.9,-1.25) {\footnotesize $M\!\!=\!\!1$};
			\node at (0.9,-1.5) {\footnotesize $K\!\!=\!\!0$};
			\node[color=myblue] at (0.9,-1.75) {\footnotesize $C\!\!=\!\!1$};
			\draw[color=myblue,thick] (0.6,0.4) -- (1.2,0.4);
			\node at (1.5,-1.25) {\footnotesize $M\!\!=\!\!0$};
			\node at (1.5,-1.5) {\footnotesize $K\!\!=\!\!1$};
			\node[color=myblue] at (1.5,-1.75) {\footnotesize $C\!\!=\!\!1$};
			\draw[color=myblue,thick] (1.2,0.3) -- (1.8,0.3);
			\node at (2.1,-1.25) {\footnotesize $M\!\!=\!\!1$};
			\node at (2.1,-1.5) {\footnotesize $K\!\!=\!\!1$};
			\node[color=myblue] at (2.1,-1.75) {\footnotesize $C\!\!=\!\!0$};
			\draw[color=myblue,thick] (1.8,-0.5) -- (2.4,-0.5);
			\draw (0,-1) -- (2.4,-1);
			\draw[->,>=stealth] (0,-1) -- (0,1);
		\end{scope}
	\end{tikzpicture}
	\caption{\label{fig:OTP} \textbf{Difference between abstract description and implementation of the OTP.} Adding up the voltages $V_M$ and $V_K$ results in a voltage $V_C$. To ensure the secrecy of the message, this voltage has to be the same regardless of how $C$ was computed (e.g., $V_0$ should not depend on whether $C=0$ was the result of adding $0$ and $0$ or $1$ and $1$.) In practice, this can never be achieved perfectly. Hence, an adversary could gain access to the key and message by simply measuring the (public) signal $V_C$. However, it is not an issue in actual implementations because data is usually stored (illustrated by the storage container {\color{myblue}\textsf{S}}) before it is sent over a communication channel, hence noise in the system changes the voltage slightly, effectively disguising the key and message values.}
\end{figure}
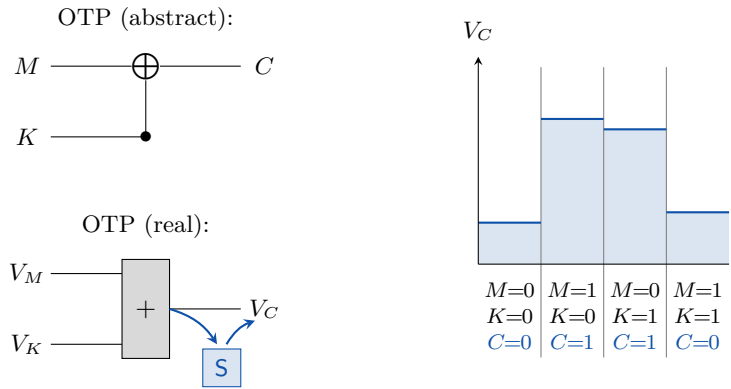

A different approach to ruling out the possibility of side-channel attacks is semi-device-inde\-pen\-dent or device-independent QKD. Here security is guaranteed from weak or even minimal assumptions about the quantum source and the detector (see, for example, \cite{Pironio2009,Braunstein2012,Lo2012}), thus narrowing the gap between protocol and implementation security. This high level of security comes, however, at a cost: In the fully device-independent case (where neither the source nor the detectors need to be characterized), the protocols require the demonstration of a loophole-free Bell test, which poses significant challenges to the experimental implementation. In 2021, the first experimental demonstrations of DIQKD have been reported \cite{Nadlinger2021,Zhang2021,Liu2022}, but the achieved parameters are still far from practical values. On a positive note, once universal quantum computers are available, they will allow for the creation of perfect Bell pairs on their logical (i.e., error-corrected) qubits.
Even though this technology is preliminary, it provides a clear path toward fully secure QKD implementations.

Classical cryptography suffers, in principle, the same threat, that is, implementations may be insecure. Indeed, side-channel attacks are a huge topic in classical cryptography and an active area of research (see \cite{Ashokkumar2016,Wu2018,Randolph2020,Panoff2022} for some examples). One example is the implementation of a \gls{OTP} (OTP) with special-purpose hardware that consists of one gate that computes the \textsf{XOR} between the key and the message (see \cref{fig:OTP}). This gate combines the voltages corresponding to the message bit $M$ and the key bit $K$, resulting in a voltage representing the ciphertext bit $C$. There are always two ways in which the value of $C$ could have been created: $C=0$ can result from adding $M=0$ and $K=0$ or $M=1$ and $K=1$. Similarly, $C=1$ can be the result of adding $M=1$ and $K=0$ or $M=0$ and $K=1$. The system's security depends on ensuring that the voltage $V_C$ representing the ciphertext bit is identical regardless of which message and key bits have been used. However, perfect equality can never be achieved in practice (as shown on the right-hand side of \cref{fig:OTP}), which means that an adversary who measures the voltage $V_C$ accurately enough can access both the message and the key values. 

The example illustrates that the problem arises in implementations where information is encoded very directly (i.e., without further stages, such as the storage of information in memory) into the state of a physical system. While this is the case in current QKD implementations, this is not a problem inherent to the use of quantum information.

\subsection{Limitation 5}
\label{lim:5}

In this last point, it is claimed that QKD increases the risk of denial of service. 

\smallskip

Classical communication networks consist of many connections, allowing for a rerouting of communication if one of these connections fails to function correctly. This redundancy helps protect them against denial-of-service attacks. Conversely, current implementations of QKD are usually based on individual point-to-point links, and an adversary with access to the link may thus easily interrupt the service. However, this is not a problem that is intrinsic to QKD. Instead, it is a consequence of the high price tag of quantum communication technology, which currently prevents us from building quantum networks with many links (see the discussion of Limitation~\hyperref[lim:2a]{2\,(a)}). In the long-term future, when larger quantum communication networks, or even a \emph{quantum internet} \cite{Wehner2018,Pompili2021,Rabbie2022,Covey2023}, are available, denial-of-service attacks can be countered by rerouting, pretty much the same way as this is done in classical networks.

\section{Outlook and recommendation}

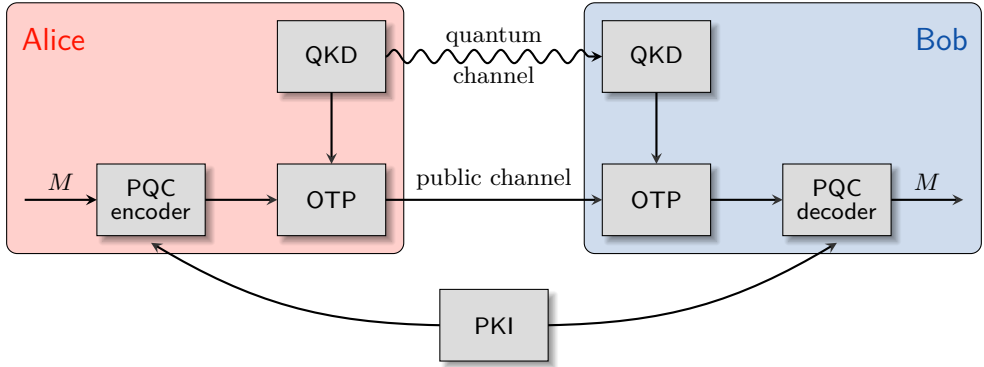
\begin{figure}[t]
	\centering
	\begin{tikzpicture}[scale=0.95]
		\draw[fill=myblue!20,rounded corners] (6.75,-0.25) rectangle (1.25,3.25);
		\draw[fill=myred!20,rounded corners] (-6.75,-0.25) rectangle (-1.25,3.25);
		\node[color=myred] at (-6.1,2.75) {\large \textsf{Alice}};
		\node at (-6,0.75) {\small $M$};
		\draw[->,>=stealth,thick] (-6.5,0.5) -- (-5.5,0.5);
		\draw[fill=LightGray,blur shadow={shadow blur steps=10}] (-3,2) -- (-1.5,2) -- (-1.5,3) -- (-3,3) -- cycle;
		\node at (-2.25,2.5) {\small \textsf{QKD}};
		\draw[->,>=stealth,thick] (-2.25,2) -- (-2.25,1);
		\draw[fill=LightGray,blur shadow={shadow blur steps=10}] (-3,0) -- (-1.5,0) -- (-1.5,1) -- (-3,1) -- cycle;
		\node at (-2.25,0.5) {\small \textsf{OTP}};
		\draw[fill=LightGray,blur shadow={shadow blur steps=10}] (-5.5,0) -- (-4,0) -- (-4,1) -- (-5.5,1) -- cycle;
		\node at (-4.75,0.65) {\small \textsf{PQC}};
		\node at (-4.75,0.35) {\small \textsf{encoder}};
		\draw[->,>=stealth,thick] (-4,0.5) -- (-3,0.5);
		\draw[->,>=stealth,decoration={snake},decorate,color=black,thick] (-1.5,2.5) to node[above] {\small quantum} node[below] {\small channel} (1.5,2.5);
		\node[color=myblue] at (6.2,2.75) {\large \textsf{Bob}};
		\node at (6,0.75) {\small $M$};
		\draw[->,>=stealth,thick] (5.5,0.5) -- (6.5,0.5);
		\draw[fill=LightGray,blur shadow={shadow blur steps=10}] (3,2) -- (1.5,2) -- (1.5,3) -- (3,3) -- cycle;
		\node at (2.25,2.5) {\small \textsf{QKD}};
		\draw[fill=LightGray,blur shadow={shadow blur steps=10}] (3,0) -- (1.5,0) -- (1.5,1) -- (3,1) -- cycle;
		\node at (2.25,0.5) {\small \textsf{OTP}};
		\draw[->,>=stealth,thick] (2.25,2) -- (2.25,1);
		\draw[fill=LightGray,blur shadow={shadow blur steps=10}] (5.5,0) -- (4,0) -- (4,1) -- (5.5,1) -- cycle;
		\draw[->,>=stealth,thick] (-1.5,0.5) to node[above] {\small public channel} (1.5,0.5);
		\node at (4.75,0.65) {\small \textsf{PQC}};
		\node at (4.75,0.35) {\small \textsf{decoder}};
		\draw[->,>=stealth,thick] (3,0.5) -- (4,0.5);
		\draw[fill=LightGray,blur shadow={shadow blur steps=10}] (-0.75,-1.75) rectangle (0.75,-0.75);
		\node at (0,-1.25) {\small \textsf{PKI}};
		\draw[->,>=stealth,thick] (-0.75,-1.25) to [bend left=15] (-4.75,-0.1);
		\draw[->,>=stealth,thick] (0.75,-1.25) to [bend right=15] (4.75,-0.1);
	\end{tikzpicture}
	\caption{\label{fig:hybridcrypto}\textbf{Hybrid QKD and PQC cryptosystem.} A message $M$ is first encrypted via a PQC scheme, which requires some (quantum-safe) public-key infrastructure (PKI) to distribute the required keys. The ciphertext is then additionally encrypted via a one-time pad (OTP), which uses keys from a QKD scheme.}
\end{figure}

The issues highlighted in \cite{NSAwhitepaper} are significant and impose severe limitations on the current usability of quantum cryptography. However, it is important to note that these limitations are not inherent to quantum cryptography but rather due to the early stage of the novel hardware required. Some of these limitations can be resolved in the medium-term future with the availability of cheaper and improved quantum technology (see Table~\ref{tab:limits}). Overcoming the remaining limitations, though, will require a long-term investment in developing quantum communication technology.

This, however, is worth the effort: Quantum cryptography has the potential to offer a true advantage over classical cryptography. Unlike traditional encryption schemes, which constantly need to be updated and strengthened to keep up with technological advancements, quantum cryptography breaks this cycle by providing protocol security that is invulnerable to all potential threats, including those posed by quantum computers. Not only do quantum cryptographic protocols gurantee secure communication during their execution, but they also offer everlasting security. Information communicated using quantum cryptography today will remain secure forever, regardless of future developments in software and hardware.

As quantum cryptography is not yet widely available, developing a strategy for securing sensitive data in the interim is essential. While standard encryption schemes such as RSA can still be used for data with a short shelf life (since universal quantum computers are not yet realized), data with a longer lifespan requires protection against ``store now, decrypt later'' attacks. Therefore, a combination of quantum key distribution (QKD) and post-quantum cryptography (PQC) in hybrid schemes currently offers the most secure approach to data encryption (this approach was, for example, explored in \cite{Dowling2020,Vyas2020,Alleaume2021}). A concrete scheme may look as follows (see \cref{fig:hybridcrypto}): A message is first encrypted using a PQC scheme, which may rely on public-key infrastructure. In addition, the resulting ciphertext is encrypted using a one-time pad, with cryptographic keys generated via QKD. This combination of PQC and QKD provides future-proof encryption resistant to attacks from both quantum and classical computers. The one-time pad ensures that the encryption remains secure even if the PQC scheme is broken in the long-term future. At the same time, the PQC encryption guarantees that even an adversary able to exploit flaws in the QKD implementation cannot read secret messages in the short or mid-term future. While this hybrid scheme requires additional infrastructure and thus still suffers from Limitation~\hyperref[lim:3a]{3\,(a)} and~\hyperref[lim:5]{5}, it remedies Limitation~\hyperref[lim:4]{4}. As such, it can be a viable interim solution for the medium-term future, when Limitation~\hyperref[lim:1]{1} to~\hyperref[lim:3]{3} are (largely) overcome (see \cref{tab:limits}).

Finally, we note that another current limitation of QKD, not mentioned in the NSA report, is its still relatively low key generation rate. While this is unproblematic if one uses QKD to replace an AES key regularly, it imposes severe limits on the communication rate if one uses it for one-time-pad encryption. However, for the same reasons as discussed in our reply to Limitation~\hyperref[lim:2a]{2\,(a)}, we expect this shortcoming to be overcome in the medium-term future.

\section*{Acknowledgements}
This work was supported by the Air Force Office of Scientific Research (AFOSR), grant No.~FA9550-19-1-0202, the QuantERA project eDICT, the National Centre of Competence in Research SwissMAP, and the ETH Zurich Quantum Center.

\bibliographystyle{halpha}
\bibliography{QKDresponseBib}

\printglossary[title=Glossary of technical terms]
\label{glossary}

\end{document}